\documentclass[pra,final,twocolumn,showpacs,showkeys]{revtex4}
\usepackage{amsmath}
\usepackage{graphicx}
\usepackage{amsfonts}
\usepackage{amssymb}
\usepackage{subfigure}
\begin{document}

\title{Multi-frequency control pulses for multi-level superconducting quantum circuits}

\author{Anne M. Forney}
\affiliation{Gettysburg College, Gettysburg, Pennsylvania 17325, USA}
\author{Steven R. Jackson}
\author{Frederick W. Strauch}
\email[Electronic address: ]{Frederick.W.Strauch@williams.edu}
\affiliation{Williams College, Williamstown, Massachusetts 01267, USA}

\date{\today}

\begin{abstract}
Superconducting quantum circuits, such as the superconducting phase qubit, have multiple quantum states that can interfere with ideal qubit operation.  The use of multiple frequency control pulses, resonant with the energy differences of the multi-state system, is theoretically explored.  An analytical method to design such control pulses is developed, using a generalization of the Floquet method to multiple frequency controls.   This method is applicable to optimizing the control of both superconducting qubits and qudits, and is found to be in excellent agreement with time-dependent numerical simulations.
  
\end{abstract} 
\maketitle

%% ALSO LOOK AT rphasequbit.txt in washboard paper!

%%%%End of Front Matter%%%%%%%%%%%%%%%%%%%%%%%%%%%%%%%%%%%%%%%%%%%%%%%%%
%%%%%%%%%%%%%%%%%%%%%%%%%%%%%%%%%%%%%%%%%%%%%%%%%%%%%%%%%%%%%%%%%%%%%%%%

\section{Introduction}

Superconducting circuits are a promising approach to building a large-scale quantum information processor.  Over the past ten years, quantum coherence times have improved by two orders magnitude, from ns to $\mu$s timescales \cite{Martinis2002,Martinis2005,Siddiqi2006,Yoshihara2006,Schreier2008,Houck2008}.  With this improvement has come increased attention to the fundamental quantum processes that arise when these circuits are controlled by microwave fields.  A recurring theme in recent experiments has been to characterize the multiple quantum levels that can be excited, in either the frequency or time domain.   On the one hand, these extra levels can interfere with ideal qubit operation.  There have been many theoretical studies of the imperfections that arise due to higher energy levels, a phenomenon called ``leakage'' \cite{Fazio99}.  On the other hand these higher levels can also be used advantageously, either to mediate quantum interactions between qubits \cite{Strauch2003} or to process quantum information with higher dimensional quantum systems called qudits \cite{Brennen2005}.  Recently, multiple levels of a superconducting phase qubit have been addressed by multi-frequency control fields to emulate a quantum spin (with spin $> 1/2$) \cite{Neeley2009}.  From either perspective, it is an important task to develop theoretical tools to model these quantum processes simply and accurately.

Most theoretical work has focused on the deviations from ideal qubit behavior during Rabi oscillations, and how these can be mitigated by pulse shaping techniques \cite{Steffen2003,Amin2006}.  Optimal control theory has also been applied to this problem \cite{Rebentrost2009,Safaei2009,Jirari09}, and recent work has indicated that arbitrarily fast control is possible using certain choices of pulses \cite{Motzoi2009}.   The presence of higher levels is also problematic for coupled qubit operation.  These arose in the study of coupled phase qubits: the spectroscopic signatures were analyzed in \cite{Johnson2003}, while a non-adiabatic controlled phase gate using the higher levels was first proposed in \cite{Strauch2003}.

Experimentally, the effect of the higher levels in a superconducting circuit has been demonstrated in transmon circuits \cite{Koch2007}, both in single-qubit operations \cite{Chow2009} and recently in a two-qubit controlled-phase gate  \cite{DiCarlo2009} similar the phase qubit gate described above.  For phase qubits, multilevel Rabi oscillations \cite{Claudon2004,Claudon2008} and multi-photon Rabi oscillations \cite{Dutta2008} have been analyzed in some detail, while a sensitive characterization of leakage was demonstrated by a Ramsey filter method \cite{Lucero2008}.  Recently, interference effects due to multiple frequency controls were demonstrated \cite{Sillanpaa2009}, realizing effects related analogous to electromagnetically-induced transparency \cite{Murali2004,Dutton2006}.

In this paper we develop a simple theoretical framework to describe the control of multiple levels in a superconducting  phase qubit using multi-frequency control fields.  We start from an early proposal to reduce leakage during qubit manipulation by resonantly cancelling off-resonant transitions to the higher energy levels \cite{Tian2000}.  Numerical simulations are used to demonstrate that this approach can optimize a quantum transition on the multi-level qubit.  These results are explained using the many-mode generalization \cite{Ho1983} of the Floquet formalism \cite{Shirley1965} for a Hamiltonian that is periodic in time.  We show that multi-frequency control fields can produce a unique quantum interference to optimize the desired transition, without complex pulse shaping.  We further show how the Floquet formalism can describe other interference effects when driving multiple transitions.

This paper is organized as follows.  In Section II we describe the basic model of a phase qubit.  In Section III, we introduce the Floquet formalism for a single frequency control pulse, reproducing the effects that occur in three-level Rabi oscillations.  This formalism generalizes the rotating wave approximation, taking a time-dependent problem to a time-independent problem (with a much larger state space).  In Section IV we extend the Floquet formalism to include control fields with multiple frequencies.  This analytical approach is used to optimize a transition between the first two levels of the phase qubit.  These ideas are confirmed in Section V through numerical optimizations of square and Gaussian control pulses.  We return to the Floquet formalism in Section VI to predict beating effects relevant to the recent spin emulation experiment  \cite{Neeley2009}.  Finally, we conclude our study in Section VII, while certain theoretical results are detailed in the Appendix.

\section{Phase Qubit Hamiltonian}

The phase qubit is generally based on a variation of the current-biased Josephson junction \cite{Martinis2002}.  This is described by the following Hamiltonian
\begin{equation}
H =4 E_c \hbar^{-2} p_{\gamma}^2 - E_J \left(\cos \gamma + (I/I_c) \gamma\right).
\end{equation}
The dynamical variables are $\gamma$, the gauge-invariant phase difference, and $p_{\gamma}$, its conjugate momentum, subject to the commutation relation $[\gamma,p_{\gamma} ] = i \hbar$.  The other parameters are the junction's bias and critical currents $I = I_{dc}$ and $I_c$, the capacitance $C$, and the energy scales $E_J = \hbar I_c / 2e$ and $E_c = e^2 / 2C$.

To describe Rabi oscillations, we will let the bias current be time-dependent, of the form $I = I_{dc} - I_{ac} (t)$, and restrict the Hamiltonian to the lowest four energy levels to find
\begin{equation}
H = H_0 + f(t) X,
\end{equation}
where we have divided the Hamiltonian into its unperturbed, time-independent form
\begin{equation}
H_0 = \left(\begin{array}{cccc} E_0 & 0 & 0 & 0\\ 0 & E_1 & 0 & 0\\ 0 & 0 & E_2 & 0 \\ 0 & 0 & 0 & E_3\end{array}\right),
\end{equation}
and a set of dimensionless matrix elements
\begin{equation}
X = \left(\begin{array}{cccc} x_{00} & x_{01} & x_{02} & x_{03} \\ x_{01} & x_{11} & x_{12} & x_{13} \\ x_{02} & x_{12} & x_{22} & x_{23} \\ x_{03} & x_{13} & x_{23} & x_{33}\end{array}\right).
\end{equation}
The energy levels $E_n$ and the matrix elements $x_{nm}$ can be calculated by either diagonalizing the washboard potential directly, or by some approximation scheme.  The latter can be efficiently performed by first approximating the washboard potential as a cubic oscillator, of the form
\begin{equation}
H = \hbar \omega_0 \left(\frac{1}{2} p^2 + \frac{1}{2} x^2 - \lambda x^3 \right),
\end{equation}
where $\hbar \omega_0 = \sqrt{8 E_c E_J} \left(1 - (I_{dc}/I_c)^2 \right)^{1/4}$ and $\lambda = 1/\sqrt{54 N_s}$ with $N_s$ given by
\begin{equation}
N_s = \frac{\Delta U}{\hbar \omega_0} \approx \frac{2^{3/4}}{3} \left(\frac{E_J}{E_c}\right)^{1/2} \left(1-\frac{I_{dc}}{I_c}\right)^{5/4}.
\end{equation}
The resulting energies and matrix elements, calculated using perturbation theory, are found in the Appendix.  Finally, the driving field has the explicit form
\begin{equation}
f(t) = \frac{I_{ac}(t)}{I_c}  E_J \left(\frac{8 E_c}{\hbar \omega_0}\right)^{1/2} \approx - \hbar \frac{d \omega_0}{d I} \frac{I_{ac}(t)} {3 \lambda}.
\end{equation}

\section{Single-mode Floquet Theory: Three-level Rabi Oscillations}
For Rabi oscillations in the presence of strong driving, there are deviations from two-level behavior that can be analyzed using a three-level model.  Previous studies \cite{Steffen2003,Meier2005,Amin2006,Strauch2007}, using the rotating wave approximation, have identified three main features.  First, the coherent oscillations between the ground and first excited state are accompanied by oscillations to the second excited state.  Second, there is a reduction in a Rabi frequency.   Finally, there is a Stark shift of the optimal resonance condition.  All of these effects have been seen experimentally \cite{Strauch2007,Lucero2008,Dutta2008}.   In this section we theoretically derive these effects by introducting the Floquet formalism \cite{Shirley1965}.  

First, we let the driving field be given by $f(t) = A \cos \omega t$.  Then, we expand the wavefunction as a Fourier series
\begin{equation}
|\Psi(t) \rangle = \sum_{n=-\infty}^{\infty} |\psi_n(t) \rangle e^{i n \omega t}.
\end{equation}
Finally, substituting this series into the Schr{\"o}dinger equation $i \hbar d |\Psi\rangle / dt  = H |\Psi\rangle$, with $H = H_0 + A X \cos \omega t$, we match terms proportional to $e^{i n \omega t}$ on each side.  The resulting equations to be solved are
\begin{equation}
i \hbar \frac{d |\psi_{n}\rangle}{dt} = (H_0 + n \hbar \omega) |\psi_{n} \rangle + \frac{1}{2} A X (|\psi_{n-1}\rangle + |\psi_{n+1}\rangle).
\end{equation}
Letting $|\psi_n(t)\rangle = e^{-i \bar{E} t/\hbar} |\psi_n(0)\rangle$, we find that these coupled equations are equivalent to a time-independent Schr{\"o}dinger equation $\mathcal{H}_{F} |\Psi\rangle = \bar{E} |\Psi\rangle$ for the infinite state $|\Psi\rangle = \sum_{n=-\infty}^{\infty} |\psi_{n}\rangle \otimes |n\rangle$ with the Floquet Hamiltonian matrix
\begin{equation}
(\mathcal{H}_{F})_{n,m} = (H_0 + n \hbar \omega) \delta_{n,m} + \frac{1}{2} A X (\delta_{n,m-1} + \delta_{n,m+1}).
\end{equation}
The labels $n$ and $m$ can be interpreted as photon numbers for the driving field, and the overall state as that of the combined system and field.

In general, this approach has replaced a finite-dimensional time-dependent problem with an infinite-dimensional time-independent problem.  To solve the latter, we can approximate the infinite matrix by one of its sub-blocks.  For the problem at hand, the lowest-order approximation is to include only three states: $|0,0\rangle$, $|1,-1\rangle$, and $|2,-2\rangle$, where we are using the notation of the form $|s,n\rangle$ to indicate the system in state $s=0, 1, 2$ with $n=0$, $-1$, and $-2$ photons, respectively.  Negative photon numbers are allowed here, as these are differences from the average photon number in a semi-classical state \cite{Shirley1965}.  After removing an overall constant energy $E_0$, the resulting Floquet matrix takes the form
\begin{equation}
\mathcal{H}_{F} =  \hbar \left(\begin{array}{ccc} 0 & \Omega_{01} /2 & 0 \\ \Omega_{01} /2 & \omega_{01}-\omega & \Omega_{12}/2 \\ 0 & \Omega_{12}/2 & \omega_{02} - 2 \omega \end{array}\right),
\end{equation}
where $\hbar \Omega_{01} = A x_{01}$,  $\hbar \omega_{01} = E_1 - E_0$,  and $\hbar \omega_{02} = E_2 - E_0$.  For convenience, we also define $\hbar \omega_{12} = E_2 - E_1$; note that $\omega_{02} = \omega_{12} + \omega_{01}$.  Note also that this approach reproduces the rotating wave approximation exactly, while including more states allows for systematic corrections due to strong multiphoton processes, such as the Bloch-Siegert shift \cite{Shirley1965}.  The resulting dynamics can be found by diagonalizing the Floquet matrix.  For this Hamiltonian exact results are available \cite{Amin2006,Strauch2007}; here we will adopt a perturbative approach.  

To simplify the following, we consider the case of near resonance with $\delta = \omega - \omega_{01} \ll \Omega_{01}$, and approximate $\Omega_{12} \approx \sqrt{2} \Omega_{01}$.  For weak driving, the largest scale in the problem is $\Delta = 2 \omega - \omega_{02} \approx \omega_{01} - \omega_{12} \approx  5 \omega_0 / (36 N_s)$.  This corresponds to the anharmonicity of the system, being inversely proportional to $N_s$.  For Rabi frequencies near this value, three-level effects become important \cite{Amin2006,Strauch2007}.  Therefore, to see deviations from two-level behavior, we use perturbation theory in the small parameters $\delta/\Omega_{01}$ and $\Omega_{01}/\Delta$, starting from the zeroth order eigenstates $(|0,0\rangle \pm |1,-1\rangle)/\sqrt{2}$ and $|2,-2\rangle$.  Using standard methods of perturbation theory, we compute the (normalized) eigenstates  $|v_{\ell}\rangle$ and eigenvalues $\bar{E}_{\ell}$, from which we calculate the time-dependent amplitudes
\begin{equation}
a_{s}(t) = \sum_{\ell=0}^2 e^{-i \bar{E}_{\ell} t/\hbar} \langle s, -s | v_{\ell} \rangle \langle v_{\ell} | \psi(0)\rangle,
\end{equation}
where we assume that $\psi(0)\rangle = |0,0\rangle$.  This calculation is best done using a computer, as the required order of perturbation theory is second order for the wavefunction and fourth order for the energy.  Alternatively, one can expand the exact eigenvalues using the roots of a cubic polynomial.  In either case, we find that the amplitudes satisfy
%\begin{equation}
%a_0(t) \approx \cos(\Omega t/2) - i \frac{\bar{\delta}}{\Omega_{01}} \sin(\Omega t/2)
%\end{equation}
\begin{equation}
a_0(t) \approx \cos(\Omega t/2) - i \sin(\Omega t/2) \left(\frac{\delta}{\Omega_{01}} - \frac{ \Omega_{01}}{2\Delta}\right),
\end{equation}

\begin{equation}
a_1(t) \approx -i \sin(\Omega t/2) \left[1 - \frac{\Omega_{01}^2}{4 \Delta^2} - \frac{1}{2}\left(\frac{\delta}{ \Omega_{01}} - \frac{ \Omega_{01}}{2\Delta}\right)^2\right],
\end{equation}
and
\begin{equation}
a_2(t) \approx -i \frac{\sqrt{2} \Omega_{01}}{2 \Delta} \sin(\Omega t/2),
\end{equation}
where the Rabi frequency $\Omega$ is given by
\begin{equation}
\Omega =  \Omega_{01} \left(1 - \frac{\Omega_{01}^2}{4 \Delta^2}\right) + \frac{\Omega_{01}}{2}\left(\frac{\delta}{ \Omega_{01}} - \frac{ \Omega_{01}}{2\Delta}\right)^2.
\end{equation}

There are many things to note about this solution.  First, we observe that both $a_1(t)$ and $a_2(t)$ are proportional to $\sin(\Omega t/2)$.  Thus, transitions from the ground to the first excited state leak out to the second excited state, with probability 
\begin{equation}
p_2(t) = |a_2(t)|^2 \approx \frac{\Omega_{01}^2}{2 \Delta^2} \sin^2(\Omega t/2).
\end{equation}
Avoiding this leakage through pulse shapes has been the subject of much investigation \cite{Steffen2003,Amin2006,Rebentrost2009,Motzoi2009}.  In addition to this error, however, is the reduction of $p_1(t) = |a_1(t)|^2$ by the factor depending on $\delta/\Omega_{01}$.  This is due to the fact that, when coupled to the second excited state, the $ 0 \to 1$ transition is no longer located at $\delta = \omega - \omega_{01} = 0$, but rather at $\delta = \Omega_{01}^2/(2\Delta)$, i.e.
\begin{equation}
\omega \approx \omega_{01} + \frac{\Omega_{01}^2}{2 (\omega_{01} - \omega_{12})}.
\end{equation}
This is the effective ac Stark shift measured in experiments \cite{Strauch2007,Dutta2008}.  As shown below, it must be compensated for high-fidelity qubit rotations.  Finally, the on-resonance Rabi frequency is given by
\begin{equation}
\Omega_{R} \approx \Omega_{01} \left(1 - \frac{\Omega_{01}^2}{4 (\omega_{01}-\omega_{12})^2}\right).
\end{equation}
Its reduction is due to the dressed eigenstates of the system, and has also been measured experimentally \cite{Claudon2004,Strauch2007,Dutta2008,Claudon2008}

\section{Two-Mode Floquet Theory: Optimized Rabi Oscillation}
In the presence of a control field of the form 
\begin{equation}
f(t)  =A_1 \cos \omega_1 t + A_2 \cos (\omega_2 t + \phi),
\end{equation}
the Floquet method can be generalized \cite{Ho1983} to include two sets of photon states for the two oscillatory components of the field.  That is, by performing the double Fourier expansion
\begin{equation}
|\Psi(t) \rangle = \sum_{\{n_1,n_2\}=-\infty}^{\infty} |\psi_{n_1,n_2}(t) \rangle e^{i n_1 \omega_1 t} e^{i n_2 \omega_2 t},
\end{equation}
the Schr{\"o}dinger equation leads to the set of coupled equations
\begin{eqnarray}
i \hbar \frac{d |\psi_{n_1,n_2}\rangle}{dt} & = & (H_0 + n_1 \hbar \omega_1 + n_2 \hbar \omega_2 ) |\psi_{n_1,n_2}\rangle \\ 
\nonumber
& & + \frac{1}{2} A_1 X (|\psi_{n_1 - 1,n_2}\rangle + |\psi_{n_1 + 1, n_2}\rangle) \\
\nonumber
& & + \frac{1}{2} A_2 X (e^{i\phi} |\psi_{n_1,n_2-1}\rangle + e^{-i \phi} |\psi_{n_1,n_2 + 1} \rangle).
\end{eqnarray}

This is equivalent to a time-independent Schr{\"o}dinger equation for the infinite state $|\Psi \rangle = \sum_{n_1,n_2} |\psi_{n_1,n_2} \rangle \otimes |n_1\rangle \otimes |n_2\rangle $ with the Floquet Hamiltonian matrix
\begin{equation}
\begin{array}{lcl}
(\mathcal{H}_{F})_{n,m} &=& (H_0 + n_1 \hbar \omega_1 + n_2 \hbar \omega_2) \delta_{n,m} \\
& & +\frac{1}{2} A_1 X (\delta_{n,m-e_1} + \delta_{n,m+e_1} )  \\
& & +\frac{1}{2} A_2 X (e^{i \phi} \delta_{n,m-e_2} + e^{-i \phi}\delta_{n,m+e_2}),
\end{array}
\end{equation}
where $n = \{n_1,n_2\}$, $m = \{m_1,m_2\}$, $e_1 = \{1,0\}$, and $e_2 = \{0,1\}$.  To obtain the state amplitudes, one sums over the intermediate photon states
\begin{equation}
a_s(t) = \sum_{n_1,n_2} e^{i (n_1 \omega_1 + n_2 \omega_2) t} \langle s,n_1,n_2 | \exp \left(-i \frac{\mathcal{H}_F t}{\hbar} \right) |\psi(0)\rangle,
\end{equation}
where in the following we will assume that $|\psi(0)\rangle = |0,0,0\rangle$.
The structure of these equations is well described elsewhere \cite{Ho1983}.  Here we make the following observations.  First, to obtain accurate numerical results, one must include several photon states in the sum---including too few results in a loss of both accuracy and unitarity.    Second, one can still use perturbation theory to obtain useful analytical results, provided one identifies the appropriate states of the combined system.  

\begin{figure}
\begin{center}
\includegraphics[width=3in]{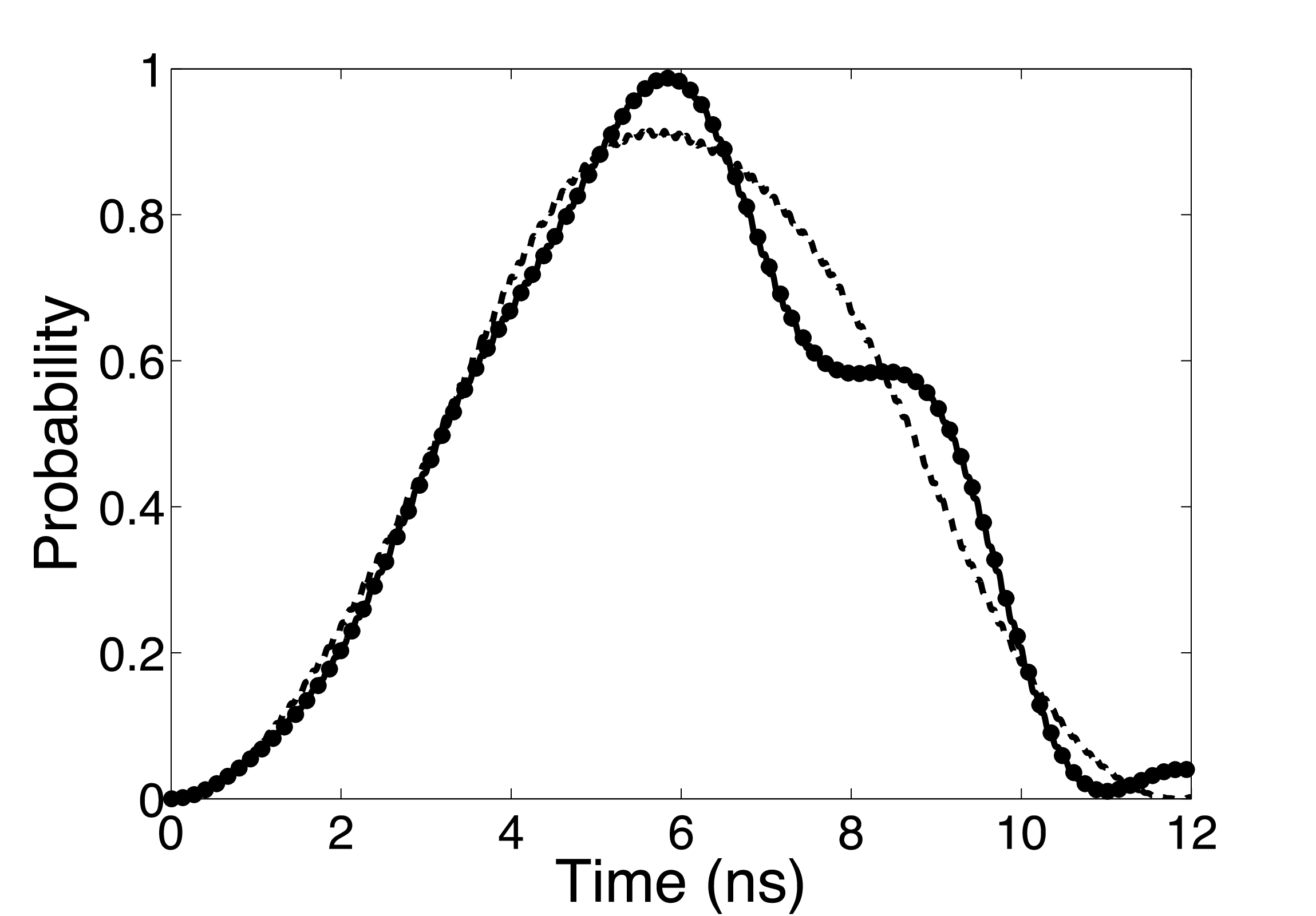}
\caption{Three-level Rabi Oscillation.  The probability $p_1(t) = |a_1(t)|^2$ to be in state 1 is shown as a function of time.  The solid curve is a numerical simulation using an optimized control field with $A_1 = 0.02 \hbar \omega_0$, $A_2 = 0.0035 \hbar \omega_0$, and $\phi = 11.44$ rad, while the dots are calculations using the two-mode Floquet formalism.  The dashed curve is a numerical simulation with $A_2 = 0$. Here, the system parameters were chosen to be $\omega_{0}/(2\pi) = 6$ GHz and $N_s = 4$.  Other relevant parameters are  $\Omega_1/(2\pi) = 86$ MHz, $\Omega_2/(2\pi) = 15$ MHz,  $\omega_1/(2\pi) = 5.785$ GHz, $\omega_{01}/(2\pi) = 5.77$ GHz, and $\omega_2/(2\pi) = \omega_{12}/(2\pi) = 5.5$ GHz.  }
\label{rabi1}
\end{center}
\end{figure}

To illustrate this method, we consider a particular example.   Fig. \ref{rabi1} shows the result of a numerical simulation of the time-dependent Schr{\"o}dinger equation for a phase qubit with $\omega_0/(2\pi) = 6$ GHz and $N_s = 4$ subject to a control field with $A_1 = 0.02 \hbar \omega_0$, $A_2 = 0.0035 \hbar \omega_0$, $\phi = 11.44$, $\omega_1 = \omega_{01} + \Omega_{01}^2/(2(\omega_{01}-\omega_{12}))$, and $\omega_2 = \omega_{12}$.  The values of $A_2$ and $\phi$ were found by a numerical search to optimize the $0 \to 1$ transition, providing a significant improvement over the $A_2=0$ dynamics.  This search was inspired by the general arguments given in Ref. \cite{Tian2000}, and demonstrates that the use of two frequencies can improve the control of this quantum system.  

Also shown in Fig. \ref{rabi1} is the result of a Floquet calculation performed by numerically diagonalizing $\mathcal{H}_F$ in a basis of $22$ states, including up to three photons for each frequency.  Here we provide an analytical approximation to explain this improved transition.  Simulations suggest that a minimal model for this transition involves the states $|0,0,0\rangle$, $|1,-1,0\rangle$, $|1,0,-1\rangle$, $|2,-2,0\rangle$, and $|2,-1,-1\rangle$.  The Floquet Hamiltonian, in this basis, reads
\begin{widetext}
\begin{equation}
\mathcal{H}_{F} = \hbar \left(\begin{array}{ccccc} 0 & \Omega_{1} /2 & \Omega_{2} e^{i \phi}/2 & 0 & 0 \\ \Omega_{1} /2 & -\delta & 0 & \Omega_{1}/\sqrt{2} & \Omega_{2} e^{i \phi}/{\sqrt{2}} \\
\Omega_{2} e^{-i \phi}/2 & 0 & \Delta & 0 & \Omega_{1}/\sqrt{2} \\ 0 & \Omega_{1}/\sqrt{2} & 0 & -\Delta_2 & 0 \\
0 & \Omega_{2} e^{-i \phi}/\sqrt{2} & \Omega_1/\sqrt{2} & 0 & -\delta
\end{array} \right),
\end{equation}
\end{widetext}
where $\hbar \Omega_1 = A_1 x_{01}$, $\hbar \Omega_{2} = A_2 x_{01}$, $\delta = \omega_1 - \omega_{01}$, $\Delta = \omega_{01} - \omega_2$, $\Delta_2 = 2 \omega_1 - \omega_{02}$, and we have let $x_{12} = \sqrt{2} x_{01}$.  By carefully normalizing and expanding out the terms found through perturbation theory, we find
\begin{equation}
a_0(t) \approx \cos(\Omega_1 t/2) \left(1 - \frac{\Omega_2^2}{\Omega_1^2}(1+ \cos(2\phi))\right) + 2 \frac{\Omega_2^2}{\Omega_1^2} e^{i 2 \delta t},
\end{equation}
\begin{eqnarray}
\nonumber
a_1(t) &\approx& - i \sin(\Omega_1 t/2) \left(1 - \frac{\Omega_1^2}{4 \Delta^2} - \frac{\Omega_2^2}{\Omega_1^2} \cos(2\phi)\right) \\ 
& & + \frac{\Omega_2}{2 \Delta}e^{-i\phi}\left(1 + 2 e^{i (\Delta + 3 \delta)t} - 3 e^{i (\Delta+ \delta) t}\right),
\end{eqnarray}
and
\begin{eqnarray}
\nonumber
a_2(t) &\approx& -i \sin(\Omega_1 t/2) \frac{\sqrt{2} \Omega_{1}}{2 \Delta} \left(1 - \frac{5 \Omega_2}{2\Omega_1} e^{-i \phi} e^{i (\Delta + \delta)t} \right) \\ \nonumber
& & -\cos (\Omega_1 t/2) \left( \frac{3 \Omega_1^2}{4 \sqrt{2} \Delta^2} - \frac{\sqrt{2} \Omega_2}{\Omega_1} e^{-i \phi} e^{i (\Delta + \delta)t} \right) \\ 
& & - \frac{\sqrt{2} \Omega_2}{\Omega_1} e^{-i \phi} e^{i(\Delta + 3 \delta)t} + \frac{\Omega_1^2}{2\sqrt{2} \Delta} e^{i (\Delta + 3 \delta)t}.
\end{eqnarray}
We see that, in addition to the Rabi oscillation terms seen previously, there are terms that oscillate at the frequencies $\delta = \omega_1 - \omega_{01} = \Omega_1^2/(2\Delta)$ and $\Delta = \omega_{01} - \omega_{12}$.  The former oscillations are slow, and can typically be ignored, but the latter oscillations become important near the peaks of the Rabi oscillations.  One can, in fact, use this to optimize the transition.

At time $T = \pi/\Omega_1$, many terms drop out of these amplitudes, and by looking at the leading order terms of $a_2$, one finds that it will vanish provided
\begin{equation}
\Omega_2 e^{-i \phi} = \frac{\Omega_1^2}{2 \Delta} e^{-i \pi/2} e^{-i (\Delta + 3 \delta) T}.
\label{optfield}
\end{equation}
This condition, in turn, specifies the optimal amplitude and the phase of the second microwave drive.  Thus, we have identified a procedure to optimize the $0 \to 1$ transition by a controlled interference through the Floquet state dynamics.  Using this value for $\phi$ and $\Omega_2$, we find that the residual error scales as $\Omega_1^4/\Delta^4$, much better than the $\Omega_1^2/\Delta^2$ scaling found for a single frequency transition.

\section{Numerical Optimization}

The analysis of the preceding section was motivated by optimizing numerically the amplitude and phase of the second frequency for the $0\to1$ transition.  As shown above, it was found that by choosing the amplitude and phase appropriately, one can obtain significant improvement in the transition probability using control fields with constant amplitude, called square pulses.  Here we compare the analytical results with the numerically optimized parameters, and show how this approach can be used to generate optimized Gaussian pulses \cite{Steffen2003}.

\begin{figure}
\begin{center}
\includegraphics[width=3in]{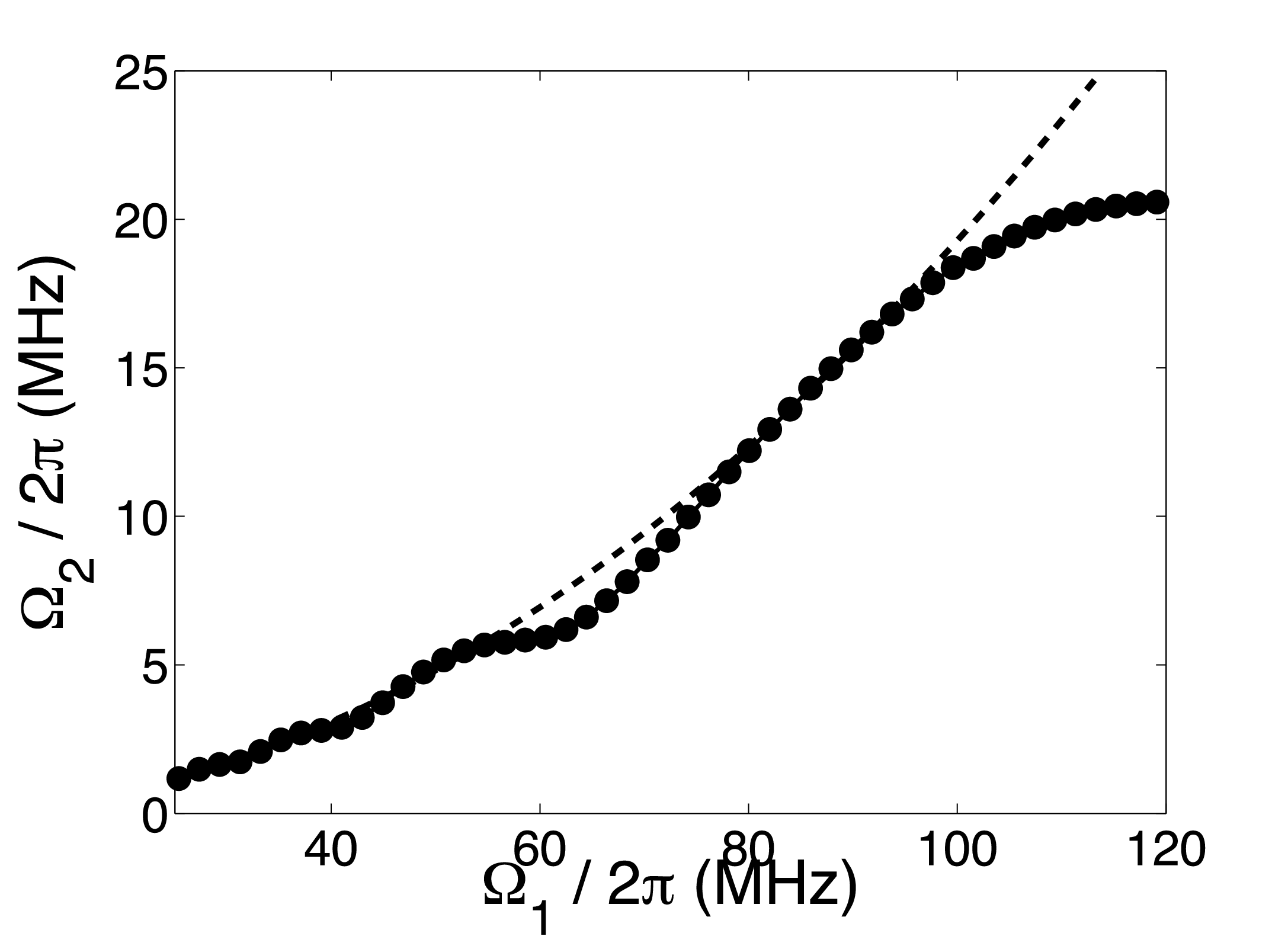}
\caption{Numerically optimized $\Omega_2$ as a function of the primary Rabi frequency $\Omega_1$.  The dashed curve is the approximation $\Omega_2 \approx \Omega_1^2/(2\Delta)$ (see text).}
\label{rabi2}
\end{center}
\end{figure}

\begin{figure}
\begin{center}
\includegraphics[width=3in]{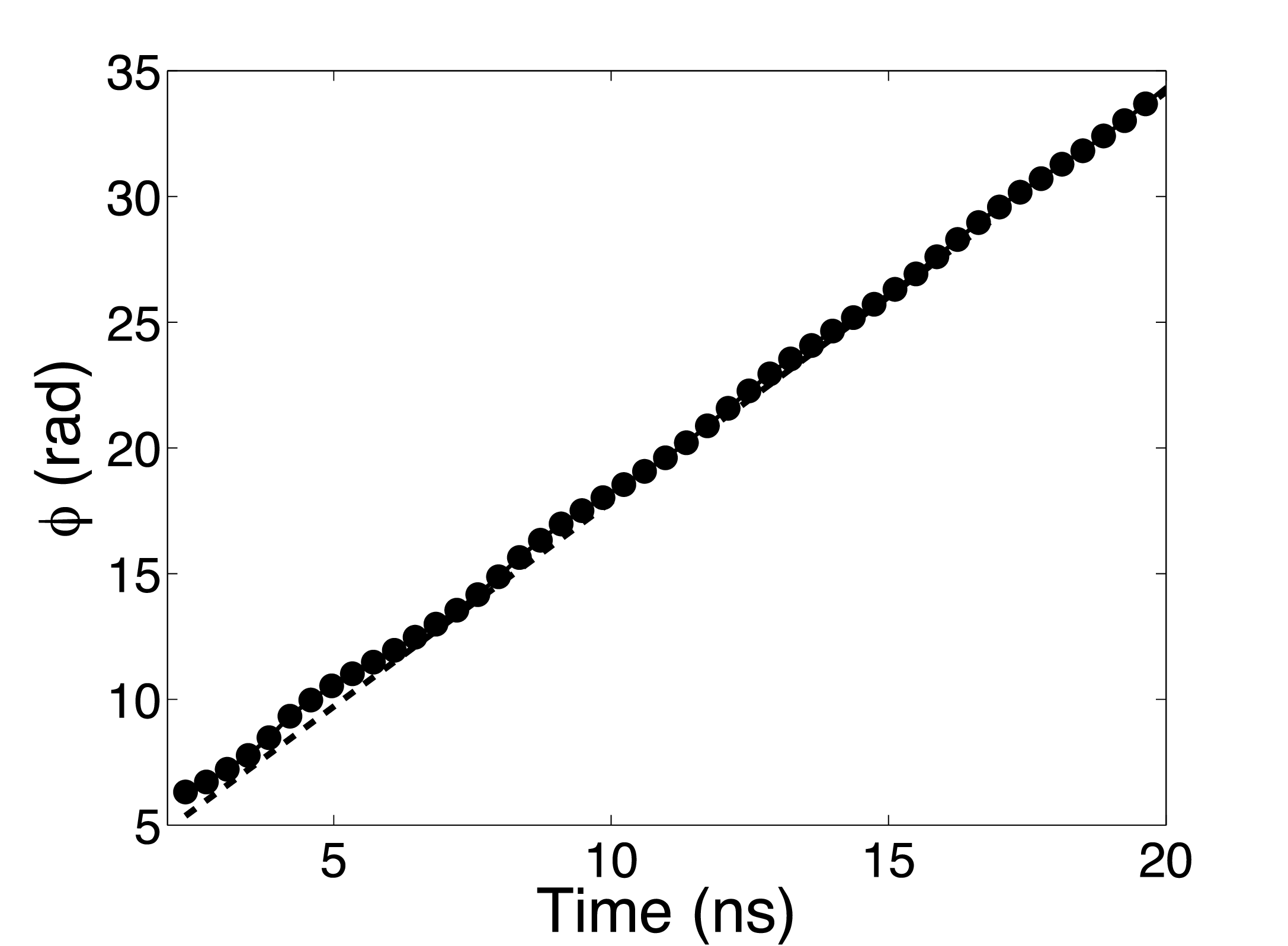}
\caption{Numerically optimized phase as a function of overall pulse time $T$.  The dashed curve is the approximation $\phi \approx \pi/2 + \Delta T$ (see text).}
\label{rabi3}
\end{center}
\end{figure}

\begin{figure}
\begin{center}
\includegraphics[width=3in]{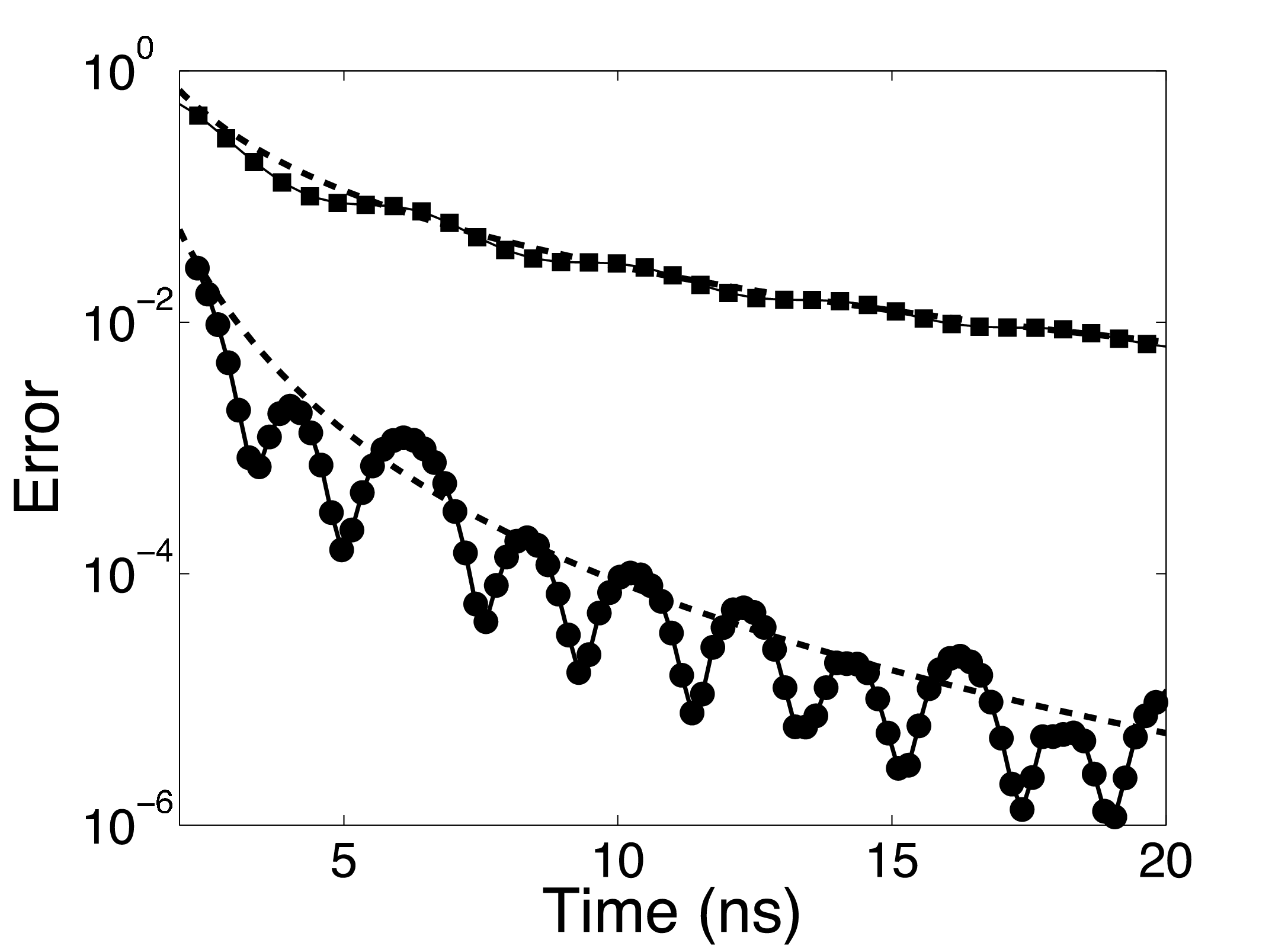}
\caption{Error of $0 \to 1$ transition using square pulses.  The upper points (squares) are the error of a pulse using a single frequency with $\omega_1 = \omega_{01}$.  The lower points (dots) are the error of an optimized two-frequency pulse with $\omega_1 = \omega_{01} + \Omega_1^2/ (2 \Delta)$ and $\omega_2 = \omega_{12}$.  The upper dashed curve is $3\Omega^2/(4 \Delta^2)$, while the lower dashed curve is $\Omega^4/(16 \Delta^4)$ (see text). }
\label{error1}
\end{center}
\end{figure}

\begin{figure}
\begin{center}
\includegraphics[width=3in]{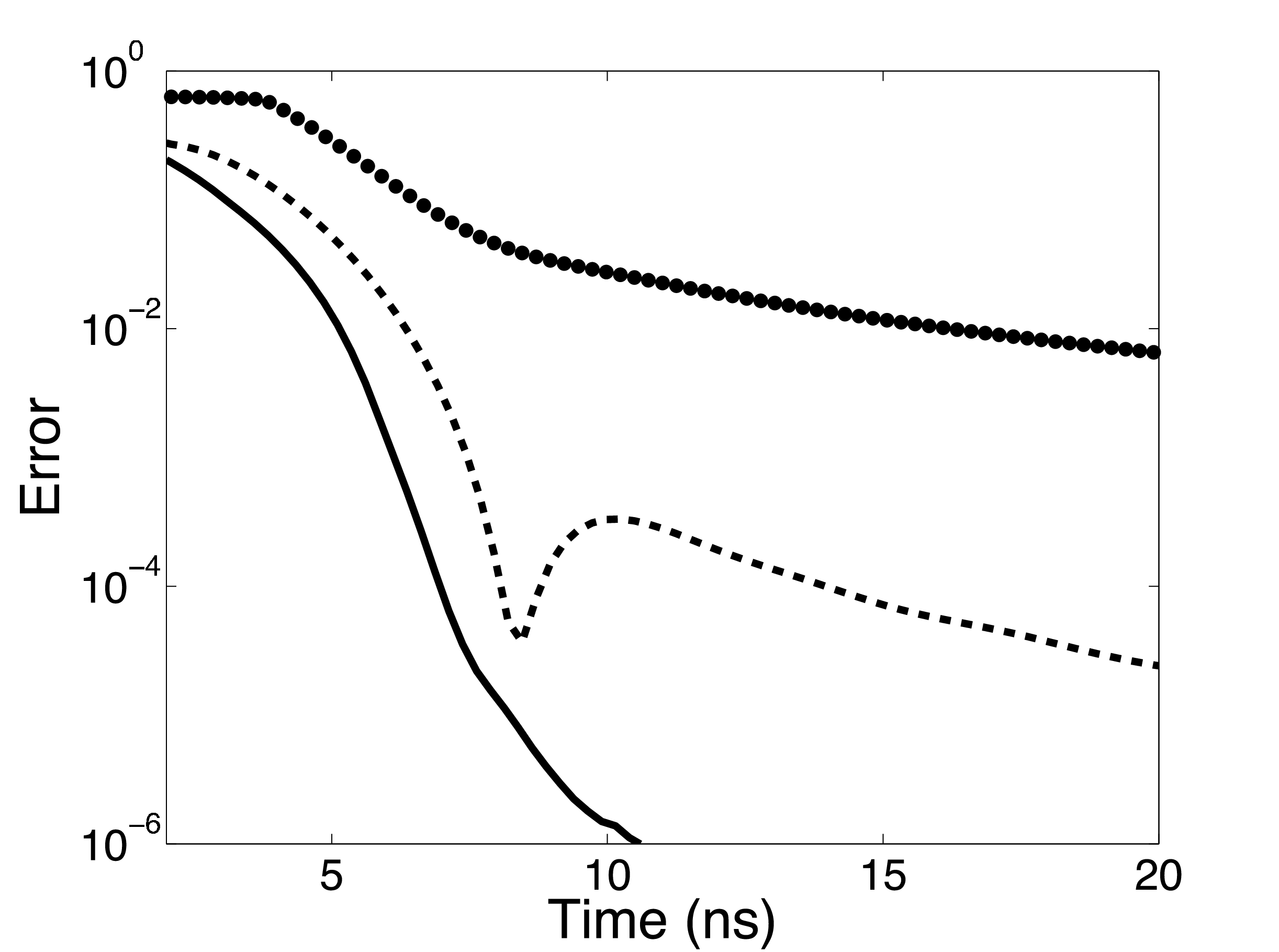}
\caption{Error of $0 \to 1$ transition using Gaussian pulses.  The upper dotted curve is the error of a pulse using a single frequency with $\omega_1 = \omega_{01}$.  The dashed curve is the error of a single-frequency pulse with $\omega_1 = \omega_{01} + d_2 \Omega_1^2/ (2 \Delta)$.  The lower solid curve is the error of an optimized two-frequency pulse with $\omega_2 = \omega_{12}$ (see text). }
\label{error2}
\end{center}
\end{figure}

First, in Fig. \ref{rabi2}, we show the numerically optimized $\Omega_2 = A_2 x_{01} /\hbar$ as a function of the bare Rabi frequency $\Omega_1 = A_1 x_{01}/\hbar$ for a phase qubit with $\omega_0/(2\pi) = 6$ GHz and $N_s = 4$, comparable to recent experiments \cite{Neeley2009}; other parameters can be found in Fig. \ref{rabi1}.  For this system, the anharmonicity is $\Delta / 2\pi \approx 260$ MHz.  We see that the analytical result
\begin{equation}
\Omega_{2,\text{opt}} \approx \frac{\Omega_1^2}{2\Delta}
\end{equation}
provides an excellent approximation for the optimized amplitude.  Similarly, the optimized phase $\phi$ is plotted as a function of the pulse time $T$ in Fig. \ref{rabi3}.  As with the amplitude, the analytical result
\begin{equation}
\phi_{\text{opt}} \approx \frac{\pi}{2} + \Delta T
\end{equation}
provides an excellent approximation.

As a further test of this method, we compare the error $p_E = 1 - p_1(T)$ for this two-frequency pulse with that of a single-frequency pulse.  This is displayed in Fig. \ref{error1}.   The single-frequency pulse is seen to have an error that scales as $\Omega_1^2/\Delta^2$.  We see that the two-frequency pulse does a significantly better job compared to the single frequency pulse, and the error scales as $\Omega_1^4/16\Delta^4$, with oscillations of frequency $\Delta$.  

Finally, using this approach, one can design pulse shapes to further optimized the transition.  We consider a Gaussian pulse shape
\begin{equation}
f(t) = s(t) \left(A_1 \cos (\omega_1 t) + A_2 \cos (\omega_2 t + \phi) \right)
\end{equation}
with
\begin{equation}
s(t) = N_{\alpha} \left(e^{- \alpha (1-2 t/T)^2} - e^{-\alpha} \right), 
\end{equation}
where $\alpha$ specifies the shape of the pulse and $N_{\alpha}$ is chosen such that $\int s(t) dt = T$ \cite{Steffen2003,Motzoi2009}.  These pulses are optimized using the bare Rabi frequency
\begin{equation}
\Omega_1 = \frac{\pi}{T} \left( 1 + c_{\alpha} \frac{ \pi^2 }{ (\Delta T)^2 } \right)
\end{equation}
and drive frequency
\begin{equation}
\omega_1 = \omega_{01}  + d_{\alpha} \frac{\pi^2}{ \Delta T^2},
\end{equation}
where the dimensionless coefficients $c_{\alpha}$ and $d_{\alpha}$ are varied to obtain the best transition. These coefficients correct for the reduction in Rabi frequency and the ac Stark shift discussed previously, and depend on the pulse shape parameter $\alpha$.  For $N_s = 4$ and $\alpha = 2$, we find that $c_{\alpha=2} = 0.58$ and $d_{\alpha = 2} = 1.245$ are required.  The error using Gaussian pulses with and without the Stark shift correction is displayed in Fig. \ref{error2}.  We see that the single-frequency pulse is not effective without these corrections.  To incorporate the two-frequency pulse, we numerically optimize for $A_2$ and $\phi$, and find that it provides a significant advantage.  Note, however, that the two-frequency square pulse outperforms all of the Gaussian pulses for small pulse times.  

\section{Three-State Oscillations}
Recently, multi-frequency control of multiple levels of a superconducting circuit has been experimentally demonstrated \cite{Neeley2009}.  This phase qudit was used to emulate spin-1 and spin-3/2 quantum systems.  Here we look at the spin-1 case, and show how the two-mode Floquet theory explains the nature of the three-state oscillations at high microwave power.

\begin{figure}
\begin{center}
\includegraphics[width=3in]{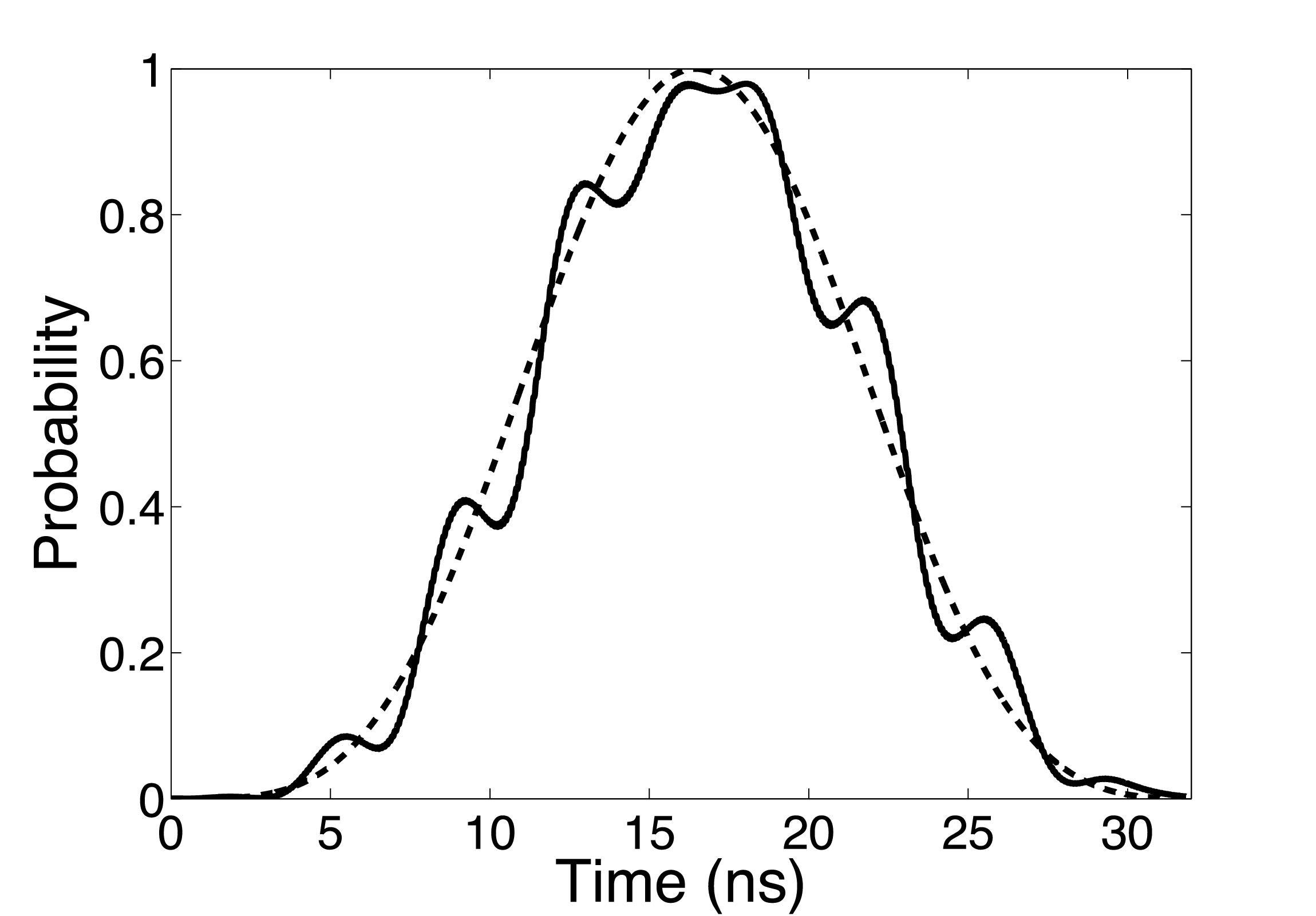}
\caption{Effective spin rotation from $0 \to 2$.  The probability $p_2(t)$ to be in state 2 is shown as a function of time.  The solid curve is a numerical simulation using $A_1 = 0.01 \hbar \omega_0$ and $A_2 = A_1 x_{01}/x_{12}$ with parameters of Fig. \ref{rabi1}.  The dashed curve is the expected rotation obtained by using the rotating wave approximation.}
\label{spin}
\end{center}
\end{figure}

Figure \ref{spin} shows the state 2 probability $p_2(t)$, when the control field is chosen with $A_1 = 0.01 \hbar \omega_0$, $\omega_1 = \omega_{01}$, $\omega_2 = \omega_{12}$, and $\hbar \Omega = A_1 x_{01} = A_2 x_{12}$.  Using the rotating wave approximation, one expects the dynamics to should emulate the rotation of a spin-1 system, yielding a probability to be in state 2 of 
\begin{equation}
p_2(t) = \sin^4 \left( \frac{\Omega t}{2 \sqrt{2}} \right).
\end{equation} 
While it is expected that there may be Stark shifts, corrections to the Rabi frequencies, and off-resonant transitions for this square pulse, a qualitatively new effect is seen in the numerical simulation.  This is a beating at the frequency $\Delta = \omega_1 - \omega_2$.

Using the Floquet formalism, one finds that the dominant effect is a coupling between three photon blocks of the three-level system, or a total of nine states:  $|0,-1,1\rangle$, $|1,-2,1\rangle$, $|2,-2,0\rangle$, $|0,0,0\rangle$, $|1,-1,0\rangle$, $|2,-1,-1\rangle$, $|0,1,-1\rangle$, $|1,0,-1\rangle$, and $|2,0,-2\rangle$.  In this basis, the effective Hamiltonian is
\begin{widetext}
\begin{equation}
\mathcal{H}_{F} = \hbar \left(\begin{array}{ccccccccc} -\Delta & \Omega /2 & 0 & 0 & \frac{1}{2} \Omega/\sqrt{2} & 0 & 0 & 0 & 0 \\
\Omega/2 &- \Delta &\Omega/2 & 0 & 0 & 0 & 0 & 0 & 0 \\
0 & \Omega/2 & -\Delta & 0 & \Omega/\sqrt{2} & 0 & 0 & 0 & 0 \\
0 & 0 & 0 & 0 & \Omega/2 & 0 & 0 & \frac{1}{2}\Omega/\sqrt{2} & 0 \\
\frac{1}{2}\Omega/\sqrt{2} & 0 & \Omega/\sqrt{2} & \Omega/2 & 0 & \Omega/2 & 0 & 0 & 0 \\
0 & 0 & 0 & 0 & \Omega/2 & 0 & 0 & \Omega/\sqrt{2} & 0 \\
0 & 0 & 0 & 0 & 0 & 0 & \Delta & \Omega/2 & 0 \\
0 & 0 & 0 & \frac{1}{2}\Omega/\sqrt{2} & 0 & \Omega/\sqrt{2} & \Omega/2 & \Delta & \Omega/2 \\
0 & 0 & 0 & 0 & 0 & 0 & 0 & \Omega/2 & \Delta 
\end{array} \right).
\end{equation}
\end{widetext}
Note that one 3-level block is isomorphic to a spin operator for a spin-1 system.

By performing the lowest order of perturbation theory for the coupling between blocks of $\mathcal{H}_F$, one finds that the relevant transition amplitude is
\begin{equation}
a_2(t) \approx - \sin^2\left(\frac{\Omega t}{ 2 \sqrt{2}}\right) - i \frac{\Omega}{4 \Delta} \sin\left(\frac{\Omega t}{\sqrt{2}}\right) (1 + 2 e^{-i  \Delta t}).
\end{equation}
This provides an excellent approximation to the beating observed in Fig. \ref{spin}.  Note that the perturbation, which is proportional to $\Omega/\Delta$, happens to vanish precisely when the unperturbed oscillation reaches its maximum ($t = \sqrt{2} \pi/\Omega$.)  Thus, it is likely that additional effects limit this approach to a $0 \to 2$ transition.  By extending the matrix to 15 states and higher orders in perturbation theory, one finds a state 3 population proportional to $\Omega^2/\Delta^2$.

\section{Conclusion}

In this paper we have analyzed a set of multi-level effects found in superconducting circuits such as the phase or transmon qubit when controlled by pulses with two microwave frequencies.  These involve a combination of resonant, off-resonant, and interference effects that are of importance for future qubit (or qudit) superconducting implementations of quantum information processors.  Indeed, we have demonstrated that the many-mode Floquet formalism for multiple frequencies is a useful generalization of the standard rotating wave approximation.  

First, we used the single-mode formalisim to recover compact analytical results for corrections to Rabi oscillations in a three-level system, finding corrections to both the resonance condition and the oscillation frequency of relative order $\Omega^2/\Delta^2$.  These are the ac Stark shift and reduction in Rabi frequency seen in existing experiments and predicted previously.

Second, we have shown that simultaneously controlling the qubit with two frequencies, one resonant with the $0 \to 1$ transition (after compensating for the ac Stark shift) and the other resonant with the $1 \to 2$ transition leads to a useful interference effect.  This insight was inspired by numerical results on square pulses, and found to be in excellent agreement.  This approach was further extended numerically to show that two-frequency Gaussian pulses can be developed for the $0 \to 1$ transition with significant improvements over single-frequency pulses.

Finally, we have used the Floquet method to explain off-resonant couplings that emerge when using the phase qudit to emulate a spin system.  Here we have found and explained a beating that is proportional to $\Omega/\Delta$, and should be observable in recent experiments, provided it is not masked by effects of decoherence.

In order for these effects to be genuinely useful, one would like to extend the optimization of a transition between two (or more) states to the optimization of a unitary operation acting an a superposition of these states.  Here, however, an interesting difficulty emerges.  For the single-frequency pulse, with square or Gaussian shapes, this is immediate: this control pulse is symmetric under time-reversal: $f(-t) = f(t)$.  Consequently, the transition from $0 \to 1$ and its time reverse from $1 \to 0$ are both optimized for a single $f(t)$.  For the two-frequency pulse, however, $f(-t) \ne f(t)$, and in fact the optimization developed in Sec. III does not perform as well for the $1 \to 0$ transition.  Note that this observation sheds some light on the two-quadrature approach of Ref. \cite{Motzoi2009}: the class of control pulses advocated there is time-reversal symmetric.  We expect that combining multiple quadratures and multiple frequencies will significantly expand the control techniques for future experiments.  Developing simple, accurate, control pulses for multi-level quantum systems remains a challenging problem for theory and experiment.

\appendix*
\section{}
In this appendix we summarize the perturbative results for the cubic oscillator
\begin{equation}
H = \hbar \omega_0 \left(\frac{1}{2} p^2 + \frac{1}{2} x^2 - \lambda x^3 \right).
\end{equation}Here we summarize the Rayleigh-Schr{\"o}dinger perturbation expansion for the Hamiltonian $H=H_0 + \lambda V$.  First, one expresses the $n$-th energy eigenstate, $|\Psi_n\rangle$, in powers of $\lambda$
\begin{equation}
|\Psi_n\rangle = \sum_{k=0}^{\infty} \lambda^k |n,k\rangle.
\label{rspt1}
\end{equation}
In this expansion, $|n,0\rangle$ is the $n$-th energy eigenstate of $H_0$, and $|n,k\rangle$ are the $k$-th order perturbative corrections.  We also expand the energy eigenvalue in powers of $\lambda$,
\begin{equation}
E_n = \sum_{k=0}^{\infty} \lambda^k E_{n,k},
\label{rspt2}
\end{equation}
where $H_0 |n,0\rangle = E_{n,0} |n,0\rangle$.  Substituting (\ref{rspt1}) and (\ref{rspt2}) in the eigenvalue equation 
\begin{equation}
(H_0 + \lambda V) |\Psi_n\rangle = E_n |\Psi_n\rangle,
\end{equation}
equating like powers of $\lambda^k$, and projecting onto $\langle m,0|$ allows one to solve for the energies and eigenfunctions:
\begin{equation}
E_{n,k} = \langle n,0|V|n,k-1\rangle
\label{rspt3}
\end{equation}
and
\begin{widetext}
\begin{equation}
|n,k\rangle = \sum_{m \ne n} \frac{\langle m,0|V|n,k-1\rangle - \sum_{j=1}^{k-1} E_{n,j} \langle m,0|n,k-j\rangle}{E_{n,0}-E_{m,0}} |m,0\rangle.
\label{rspt4}
\end{equation}
Extending this calculation to $\lambda^8$ one finds
\begin{eqnarray}
E_{n}/\hbar \omega_0 &=& (n+1/2) - \frac{1}{8} \lambda^2 (30 n^2 + 30 n + 11) - \frac{15}{32} \lambda^4 (94 n^3 + 141 n^2 + 109 n + 31) \notag \\
& &- \frac{1}{128} \lambda^6 (115755 n^4 + 231510 n^3 + 278160 n^2 + 162405 n + 39709) \notag \\
& & - \frac{21}{2048} \lambda^8 (2282682 n^5 + 5706706 n^4 + 9387690 n^3 + 8374830 n^2 + 4244573 n + 916705).
\label{cubicpt1}
\end{eqnarray}
\end{widetext}

This procedure was implemented in Mathematica to calculate the eigenvalues up to $\lambda^6$; the $\lambda^8$ expression (\ref{cubicpt1}) was found using a more efficient recursion-relation method \cite{Bender99}, and agrees with \cite{Alvarez89} (provided one lets  $4^N \to 4^{2N}$).   These results, when compared with numerical results found by complex scaling \cite{Yaris78L}, are found to be accurate for states $n=0\to 3$ when $N_s > 3$.

In addition to the energy levels, perturbation theory also provides expressions for the wavefunctions.  For reference we list the third-order expression:
\begin{widetext}
\begin{equation}
|\Psi_n\rangle = |n\rangle + \lambda \sum_{k=-3}^{+3} a_k(n) |n+k\rangle + \lambda^2 \sum_{k=-6}^{+6} b_k(n) |n+k\rangle + \lambda^3 \sum_{k=-9}^{+9} c_k(n) |n+k\rangle
\label{rspt5}
\end{equation}
where $|n\rangle=|n,0\rangle$ are the eigenstates of the purely harmonic oscillator Hamiltonian, and the nonzero expansion coefficients are 
\begin{eqnarray}
a_{-3}(n) &=& - \frac{1}{6 \sqrt{2}} \left(n (n-1)(n-2)\right)^{1/2} \\
a_{-1}(n) &=& - \frac{3}{2 \sqrt{2}} n^{3/2} \\
a_{+1}(n) &=& - \frac{3}{2 \sqrt{2}} (n+1)^{3/2} \\
a_{+3}(n) &=&  \frac{1}{6 \sqrt{2}} \left((n+1)(n+2)(n+3)\right)^{1/2}, 
\end{eqnarray}
\begin{eqnarray}
b_{-6}(n) &=&  \frac{1}{144} \left(n (n-1)(n-2)(n-3)(n-4)(n-5) \right)^{1/2} \\
b_{-4}(n) &=&  \frac{1}{32} \left(n (n-1)(n-2)(n-3)\right)^{1/2} (4n-3) \\
b_{-2}(n) &=&  \frac{1}{16} \left(n (n-1)\right)^{1/2} (7 n^2 - 19 n + 1) \\
b_{+2}(n) &=&  \frac{1}{16} \left((n+1)(n+2)\right)^{1/2} (7 n^2 + 33 n + 27) \\
b_{+4}(n) &=&  \frac{1}{32} \left((n+1)(n+2)(n+3)(n+4)\right)^{1/2} (4n+7) \\
b_{+6}(n) &=&  \frac{1}{144} \left((n+1)(n+2)(n+3)(n+4)(n+5)(n+6)\right)^{1/2}, 
\end{eqnarray}
and
\begin{eqnarray}
c_{-9}(n) &=& -\frac{1}{2592 \sqrt{2}} \left(n (n-1)(n-2)(n-3)(n-4)(n-5)(n-6)(n-7)(n-8)\right)^{1/2} \\
c_{-7}(n) &=& -\frac{1}{192 \sqrt{2}} \left(n (n-1)(n-2)(n-3)(n-4)(n-5)(n-6)\right)^{1/2} (2n -3) \\
c_{-5}(n) &=& -\frac{1}{960 \sqrt{2}} \left(n (n-1)(n-2)(n-3)(n-4)\right)^{1/2} (80 n^2 -305n + 164) \\
c_{-3}(n) &=& -\frac{1}{1728 \sqrt{2}}\left(n (n-1)(n-2)\right)^{1/2} (488 n^3 - 2175 n^2 + 4018 n - 825) \\
c_{-1}(n) &=& -\frac{3}{64 \sqrt{2}} n^{1/2} (20 n^4 + 81 n^3 + 326 n^2 + 81 n + 44) \\
c_{+1}(n) &=& \frac{3}{64 \sqrt{2}} (n+1)^{1/2} (20 n^4 - n^3 + 203 n^2 + 408 n + 228) \\
c_{+3}(n) &=& \frac{1}{1728 \sqrt{2}} \left((n+1)(n+2)(n+3)\right)^{1/2} (488 n^3 + 3639 n^2 + 9832 n + 7506) \\
c_{+5}(n) &=& \frac{1}{960 \sqrt{2}} \left((n+1)(n+2)(n+3)(n+4)(n+5)\right)^{1/2} (80 n^2 + 465 n + 549) \\
c_{+7}(n) &=& \frac{1}{192 \sqrt{2}} \left((n+1)(n+2)(n+3)(n+4)(n+5)(n+6)(n+7)\right)^{1/2} (2n+5) \\
c_{+9}(n) &=& \frac{1}{2592 \sqrt{2}} \left((n+1)(n+2)(n+3)(n+4)(n+5)(n+6)(n+7)(n+8)(n+9)\right)^{1/2}.
\end{eqnarray}
\end{widetext}
One application of these expressions is to calculate the (properly normalized) matrix elements of the position operator
\begin{equation}
x_{n,m} = \frac{\langle \Psi_n| x |\Psi_m \rangle}{(\langle \Psi_n | \Psi_n \rangle \langle \Psi_m | \Psi_m \rangle)^{1/2}}.
\end{equation}
Using the wavefunctions (\ref{rspt5}) and matrix elements of the previous section, we find
\begin{eqnarray}
x_{0,0} &=& \frac{3}{2} \lambda + \frac{33}{2} \lambda^3 \\
x_{0,1} &=& \frac{\sqrt{2}}{2} + \frac{11\sqrt{2}}{8} \lambda^2 \\
x_{0,2} &=& -\frac{\sqrt{2}}{2} \lambda - \frac{243\sqrt{2}}{16} \lambda^3 \\
x_{0,3} &=& \frac{3 \sqrt{3}}{8} \lambda^2 \\
x_{1,1} &=& \frac{9}{2} \lambda + \frac{213}{2} \lambda^3 \\
x_{1,2} &=& 1 + \frac{11}{2} \lambda^2 \\
x_{1,3} &=& -\frac{\sqrt{6}}{2} \lambda - \frac{405 \sqrt{6}}{16} \lambda^3 \\
x_{2,2} &=& \frac{15}{2} \lambda + \frac{573}{2} \lambda^3 \\
x_{2,3} &=& \frac{\sqrt{6}}{2} + \frac{33 \sqrt{6}}{8} \lambda^2 \\
x_{3,3} &=& \frac{21}{2} \lambda + \frac{1113}{2} \lambda^3 
\end{eqnarray}
with corrections of order $\lambda^4$.  

\bibliography{rabix}

\end{document}